\documentclass[runningheads]{llncs}
\usepackage{amsmath}
\usepackage{threeparttable}
\usepackage{mathrsfs}
\usepackage{multirow}
\usepackage{amsfonts}
\usepackage{booktabs}
\usepackage{float}
\usepackage{ulem}
\usepackage{threeparttable}
\usepackage{color}
\usepackage{ulem}
\usepackage{url}
\usepackage[T1]{fontenc}
\usepackage{graphicx}
\begin{document}
\title{Simultaneous Speech Extraction for Multiple Target Speakers under the Meeting Scenarios}
%
%
\author{Bang Zeng\inst{1,2} \and
Hongbing Suo\inst{3} \and
Yulong Wan\inst{3} \and
Ming Li\inst{1,2}\thanks{$\dagger$ Corresponding Author, E-mail: ming.li369@dukekunshan.edu.cn}}
\authorrunning{Bang Zeng, et al.}
%
\institute{School of Computer Science, Wuhan University, Wuhan, China \and
Suzhou Municipal Key Laboratory of  Multimodal Intelligent Systems, \\
Duke Kunshan University, Kunshan, China \and 
Data \& AI Engineering System, OPPO, Beijing, China}
\maketitle              
\begin{abstract}
The common target speech separation directly estimate the target source, ignoring the interrelationship between different speakers at each frame. We propose a multiple-target speech separation model (MTSS) to simultaneously extract each speaker's voice from the mixed speech rather than just optimally estimating the target source. Moreover, we propose a speaker diarization (SD) aware MTSS system (SD-MTSS), which consists of a SD module and MTSS module.  By exploiting the TSVAD decision and the estimated mask, our SD-MTSS model can extract the speech signal of each speaker concurrently in a conversational recording without additional enrollment audio in advance. Experimental results show that our MTSS model achieves 1.38dB SDR, 1.34dB SI-SDR, and 0.13 PESQ improvements over the baseline on the WSJ0-2mix-extr dataset, respectively. The SD-MTSS system makes 19.2\% relative speaker dependent character error rate (CER) reduction on the Alimeeting dataset.

\keywords{Target speech separation, Interrelationship, Speaker Diarization, Target Speaker Voice Activity Detection, SD-MTSS.}
\end{abstract}
\section{Introduction}


In the real world, noise and speaker interference can degrade the system performance of back-end speech applications. Speech separation effectively solves this problem by extracting the target speech from the mixed utterance. Early methods called blind speech separation, such as Deep Clustering (DPCL)~\cite{hershey2016deep}, Deep Attractor Network (DANet)~\cite{chen2017deep}, and Permutation Invariant Training (PIT)~\cite{yu2017permutation,2017multitalker}, can separate each source from a mixed speech. These algorithms formulated in the time-frequency domain have an upper bound on reconstructing waves~\cite{luo2018tasnet}. Recent solutions in the time-domain, such as Time-Domain Audio Source Separation (Tas-Net)~\cite{luo2018tasnet,luo2019conv} and Dual-Path RNN (DPRNN)~\cite{luo2020dual}, break through the constraints and achieve state-of-the-art performance in the separation task. Despite this, the unknown number of speakers and the global permutation problem are still two challenges for blind speech separation.

To address the above two problems, a framework called speaker extraction~\cite{ge2021multi,9414998} or target speech separation~\cite{Wang2018VoiceFilterTV,Li2020AtssNetTS} can extract a target speaker' speech from the mixed audio by utilizing an auxiliary reference speech of the target speaker. However, it is required to filter out multiple target speakers in certain tasks, e.g., meeting scenarios. The common approach is to infer the mixed speech for several times and each process is independent of the other, ignoring the interrelationship between the speech of different speakers at each frame. In addition, obtaining the reference speech of multiple target speakers in advance is difficult to achieve. Considering the aforementioned problems, repeatedly processing the mixture speech towards different target speakers separately may not be a feasible solution.

It is worth noting that speech in the meeting scenario usually has a long duration and contains both single-talker and overlapped voice segments. Thus, it is possible to use the single-talker segments as the reference speech for  participants instead of obtaining additional speech for enrollment. Speaker Diarization (SD)~\cite{wang2018speaker} technology is very suitable for this role. SD aims to slice different speaker segments in a continuous multiple speakers conversation and determine which speaker each segment belongs to. More recently, MC-TS-VAD~\cite{wang2022cross}, which selects TSVAD as the post-processing module and employs cross-channel self-attention, achieved the best result in the Multi-party Meeting Transcription Challenge (M2Met)~\cite{Yu2021M2MetTI}.

In this work, we propose the MTSS model, which is a speech extraction method for multiple target speakers. The MTSS model infers each speaker's mask simultaneously and limits their estimated masks to be sum to 1. We consider that the energies of different speakers at each frame are not independent to each other. Moreover, we propose the SD-MTSS  framework, which associates target speech separation with speaker diarization. We select the TSVAD system as the speech diarization network. Based on the decisions from TSVAD~\cite{Ding2020PersonalVS}, we can obtain each speaker's reference speech directly from the mixed audio. Then, each speaker's reference speech is fed into the MTSS module in the separation stage.

The rest of this work is organized as follows. In Section 2, we present the architecture of the proposed MTSS and SD-MTSS models. In Section 3, we report the experimental setup. In Section 4, we report the results and discussions. The conclusions are drawn in Section 5.

\begin{figure}[t!]
  \centering
  \includegraphics[width=0.55\linewidth]{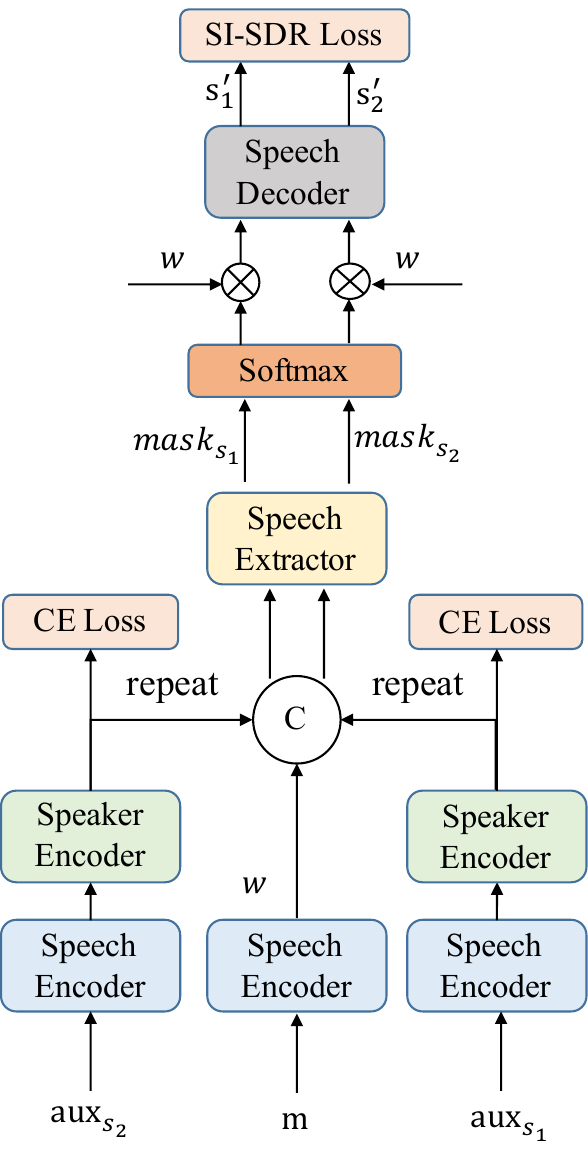}
  \caption{The details of the MTSS model. \(aux_{s_{1}}\) and \(aux_{s_{2}}\) are the reference wave of two speakers. \(m\) is the mixed wave."C" denotes the operation of concatenate.\(\otimes\) is an operation for element-wise product.}
  \label{fig:MTSS}
\end{figure}

\section{Methods}
\subsection{Multiple target speech separation model}

\subsubsection{Backbone}
  The backbone of the Multiple Target Speech Separation (MTSS) model is SpEx+~\cite{ge2020spex+}, which consists of two twin speech encoders, a speaker encoder, a speaker extractor, and a speech decoder. The twin speech encoder model the input sequence and auxiliary speech in a common latent space through sharing the structure and parameters. The speaker encoder model is a ResNet-based speaker classifier used to generate the speaker embedding of the reference speech. The speaker extractor takes both the speaker embedding and the output of the twin speech encoder as the inputs, and then produces masks in three different scales. The speech decoder outputs the estimation by multiplying the input sequence and the multi scales masks.
 
 \subsubsection{MTSS model}
  Here, we propose a speech extraction model for multiple target speakers (MTSS), which can simultaneously separate the speech of each speaker present in the conversation. The schematic diagram of the MTSS model is shown in Figure~\ref{fig:MTSS}. Unlike the original SpEx+ neural network takes only one speaker's reference speech, MTSS takes both two speaker's reference speech as the inputs and process them separatly. Moreover, we replace the ReLU with softmax to establish the relationship between the masks of each speaker in the same utterance. We believe that taking the interrelation into account will improve the final separation performance of the model. Because in the definition of binary masks, each time-frequency cell belongs to a speaker with stronger energy. Specifically, the responses of MTSS \(\boldsymbol{{s^{\prime}}_1}\), \({\boldsymbol{{s^{\prime}}_2}}\) can be formulated as:
\begin{equation}
  ({\boldsymbol{{s^{\prime}}_1}, \boldsymbol{{{s^{\prime}}_2}}}) = \boldsymbol{m} \otimes \{ \text{softmax}(cat(\boldsymbol{mask_{s_1}}, \boldsymbol{mask_{s_2}})) \}
  \label{eq1}
\end{equation}
where \(\boldsymbol{mask_{s_1}}\), \(\boldsymbol{mask_{s_2}}\) \(\in\) \(\mathbb{R}^{N \times 1 \times T}\). \(\otimes\) is an operation of element-wise product. \(softmax(*)\) and \(cat(*)\) indicates that a softmax function and concatenation operates on the penultimate dimension, respectively. \(N\) is the feature dimention and \(T\) is the time length. We also implement a multi-task learning framewrok for the target speech separation. 



\subsection{SD-MTSS system}
\begin{figure*}[t]
  \centering
  \includegraphics[width=1.0\linewidth]{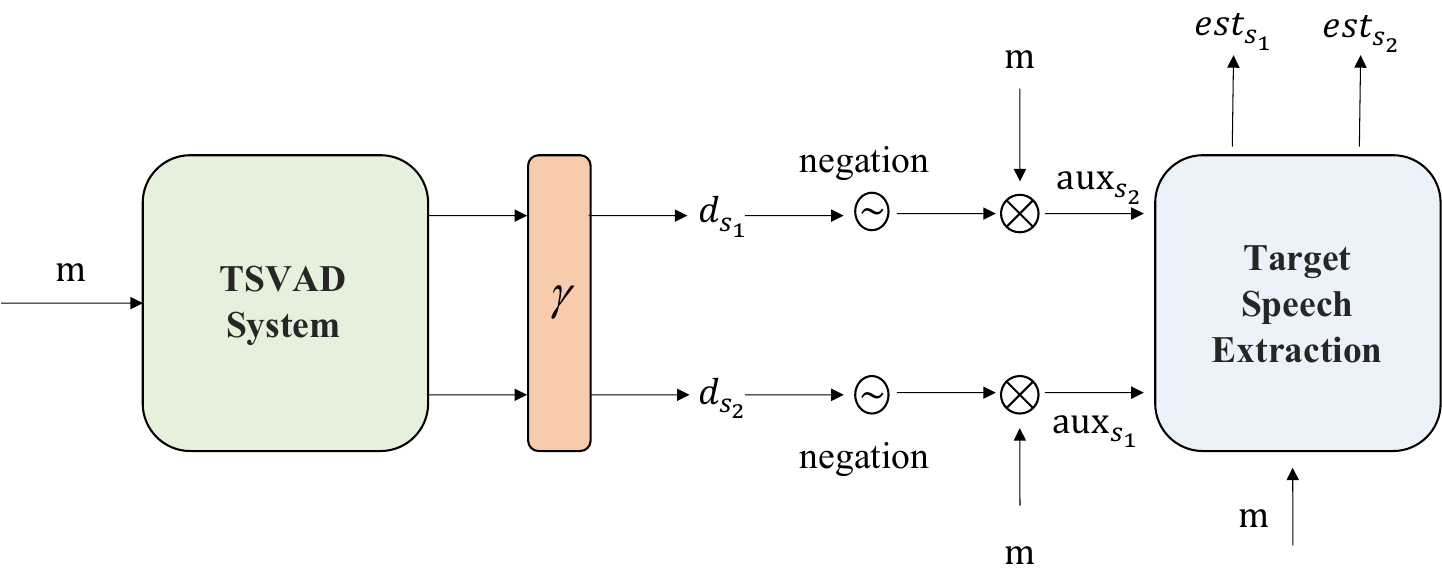}
  \caption{Schematic of the SD-MTSS system. \(\gamma\) is a threshold (0.5 often). \(m\) is the mixed input. \(s_1\), \(s_2\) represent the two speakers presented in the mixture. \(aux_{s_1}\), \(aux_{s_2}\) denote the reference speech of two speakers. \(est_{s_1}\), \(est_{s_2}\) denote the estimations of two speakers. \(d_{s_1}\), \(d_{s_2}\)indicates the binarized TSVAD decision. }
  \label{fig:system}
\end{figure*}
  Considering that it is feaible to apply speaker diarazation techniques to target speech separation, we expend the MTSS to speaker diarization (SD) aware MTSS (SD-MTSS) system. The SD-MTSS system architecture is shown in Figure~\ref{fig:system}. Rather than requiring additional registration, SD-MTSS directly obtains reference speech from the long utterance itself through the SD module. In real applications, the single-channel SD approach~\cite{wang2022} can be used here. The SD-MTSS system consists of a SD module and a MTSS module. The SD module produces the TSVAD decision for multiple speakers, which are the probabilities of each speaker's presence at the frame level. The MTSS module adopts each speaker's reference speech from the SD module and the mixture audio as inputs, and then outputs the estimation for multiple target speakers.
  
  Using the TSVAD decision, we can get the single-talker audio segments as the reference speech for each speaker. The scheme of obtaining single-talker segments in Figure~\ref{fig:system} is organized as follows. We use \(\boldsymbol{m}\) \(\in\) \(\mathbb{R}^{1 \times T}\) indicate the input sequence. \(s_1\) and \(s_2\) indicate the two different speakers in the mixture. First, the TSVAD decision is passed through a threshold mechanism and produces the binarized results \(\boldsymbol{d_{s_1}}\) and \(\boldsymbol{d_{s_2}}\) . Its values consist of 0 and 1. Then the reference speech can be formulated as:



\begin{equation}
  \boldsymbol{aux_{s_1}} = \boldsymbol{m} \otimes {\boldsymbol{\tilde{d_{s_1}}}}
  \label{eq3}
\end{equation}
\begin{equation}
  \boldsymbol{\tilde{d_{s_1}}} = (\boldsymbol{d_{s_1}} - \boldsymbol{d_{s_1} \otimes \boldsymbol{d_{s_2}}})
  \label{eqds1}
\end{equation}
\begin{equation}
  \boldsymbol{aux_{s_2}} = \boldsymbol{m} \otimes {\boldsymbol{\tilde{d_{s_2}}}}  
  \label{eq4}
\end{equation}
\begin{equation}
  \boldsymbol{\tilde{d_{s_2}}} = (\boldsymbol{d_{s_2}} - \boldsymbol{d_{s_1} \otimes \boldsymbol{d_{s_2}}})
  \label{eqds1}
\end{equation}

  where \({\boldsymbol{\tilde{d_{s_1}}}}\) and \({\boldsymbol{\tilde{d_{s_2}}}}\) indicates the mono-speaker activity part of \(\boldsymbol{d_{s_1}}\) and \(\boldsymbol{d_{s_2}}\), respectively. \(d_{s_1}\), \(d_{s_2}\) \(\in\) \(\mathbb{R}^{1 \times T}\), \(\otimes\) indicates the element-wise product. \(\otimes\) is an operation of element-wise product. Selected continuous audio segments of \(\boldsymbol{aux_{s_1}}\) and \(\boldsymbol{aux_{s_2}}\) will be fed into the MTSS module as the reference speech for the subsequent separation task.

\subsection{Speaker diarization (SD) module}

\begin{figure}[t!]
  \centering
  \includegraphics[width=1.0\linewidth]{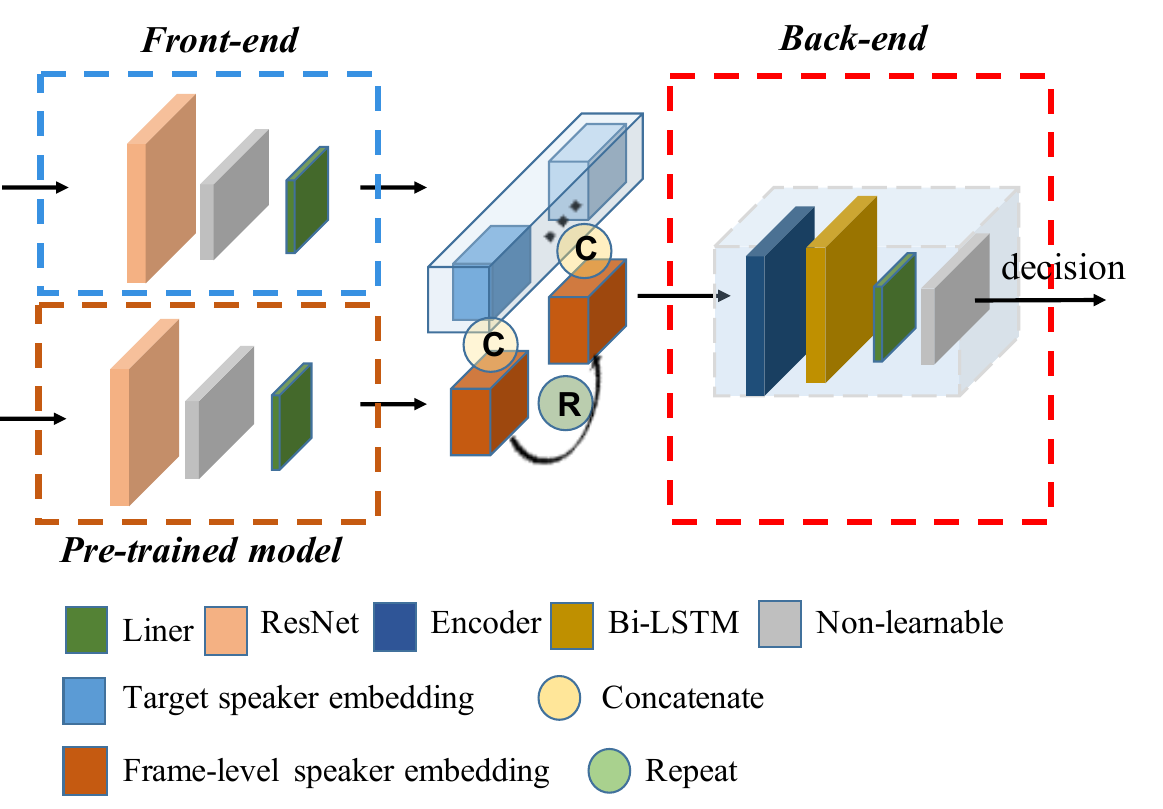}
  \caption{The structure of the TSVAD system. The Front-end share the same architecture with the pre-trained speaker embedding model. The target speaker embedding concatenates with the frame-level speaker embedding repeatedly and then is fed into the Back-end.}
  \label{fig:TSVAD}
\end{figure}

  The SD module in this work consists of a clustering-based module for target speaker embedding extraction and a TSVAD system for diarization results refinement~\cite{wang2022cross}. 
  
\subsubsection{Clustering-based module}
  The affinity matrix extraction model of TSVAD is based on the neural network in~\cite{Lin2019LSTMBS}, using an LSTM-based model in similarity measurement for speaker diarization. It consists of two bidirectional long short-term memory networks (Bi-LSTM) layers and two fully connected layers. The LSTM-based model first splits the entire audio into short speech clips and extracts the speaker embedding of all segments. Then it takes these segments as inputs and produces the initialized diarization result through adopting spectral clustering. 

\subsubsection{TSVAD system}
  The architecture of the TSVAD~\cite{wang2022cross} system is shown in Figure~\ref{fig:TSVAD}, which consists of three parts:
  \begin{enumerate}
      \item A pre-trained speaker embedding model ResNet~\cite{he2016deep} based on ArcFace~\cite{deng2019arcface} and cosine similarity scoring. The dimension of the speaker embedding layer is 128 and the margin and softmax prescaling of the ArcFace is 0.2 and 32.
      \item A front-end model with the same architecture as the pre-trained model is used to extract the frame-level speaker embedding.
      \item A back-end model consists of an encoder layer, a BiLSTM layer, a liner layer, and a sigmoid function.
\end{enumerate}

First, the pre-trained speaker embedding model extracts the target speaker embeddings. Meanwhile, the front-end network loads its parameters to extract the frame-level speaker embeddings. The target speaker embeddings are repeatedly concatenated with the frame-level speaker embeddings and then fed into the back-end. Next, the encoder layer of the back-end model produces each target speaker's detection state. The BiLSTM layer inputs these detection states and models the relationship between speakers. Finally, the linear layer coupled with a sigmoid function generates each speaker's final decision (TSVAD decision). More details can be found in~\cite{wang2022cross}.

\section{Experiment Setup}

\subsection{Dataset}
\textbf{Datasets for MTSS model:} We simulated a commonly used two-speakers mixture datasets WSJ0-2mix-extr\footnotemark[1]\footnotetext[1]{\url{https://github.com/xuchenglin28/speaker_extraction}}, the sampling rate of which is 8KHz. The simulation process is the same as~\cite{ge2020spex+}, and the only difference is that we produce a couple of target speakers speech (\(s_1\), \(s_2\)) and reference speech (\(aux_{s_1}\), \(aux_{s_2}\) ) for each mixture utterance, while~\cite{ge2020spex+} only select the first talker as the target speaker. The utterance from \(s_1\) and \(s_2\) are set in a relative SNR between 0 to 5 dB. The average SI-SDR of mixed speech is 2.50dB and -2.50dB when it takes \(s_1\) and \(s_2\) as the reference.

\noindent\textbf{Datasets for SD-MTSS model:} For the SD module, we use the training set of Alimeeting~\cite{Yu2021M2MetTI} to train the clustering-based affinity matrix extraction neural network. Alimeeting contains 118.75 hours of speech data, including 104.75 hours (426 speakers) of the training set, 4 hours (25 speakers) of the validation set, and 10 hours of the test set. For the TSVAD model in the SD module, we create a simulated datesets based on the Alimeeting training set. The simulation scheme is the same as~\cite{wang2022cross}. For MTSS module, We use the Libri-2mix~\cite{cosentino2020librimix} as the train set, its sampling rate is 16KHz. We select the siganl channel signal on channel 0 of the two-speakers samples from the Eval-Ali-far and Test-Ali-far subsets of Alimeeting to evaluate the performance of SD-MTSS model.

\subsection{Implementation details}
To compare with the baseline, the hyperparameters and learning schedule of MTSS module are set the same as \cite{ge2020spex+}. The number of filters in the encoder is 256, the number of convolutional blocks in each repeat is 8, the number of repeat is 4, the number of channels in the convolutional blocks is 512, the kernel size of the convolutional blocks is 3. The hyperparameters of the network are shown in Table~\ref{tab:hpy}.

\begin{table}[]
\caption{Hyperparameters of the MTSS module. }
  \label{tab:hpy}
  \centering
\begin{tabular}{ccc}
\hline
\textbf{Symbol} & \textbf{Settings} & \textbf{Description}                                                       \\ \hline
L1, L2, L3      & 20, 80, 160       & Lengths of the Encoder filter                                              \\
N               & 256               & Number of filters in Encoder                                      \\
X               & 8                 & Number of convolutional blocks                                             \\
B               & 256               & Number of channels in bottleneck conv blocks \\
H               & 512               & Number of channels in convolutional blocks                                 \\
P               & 3                 & Kernel size in convolutional blocks                                        \\
spk\_emb\_dim   & 256               & Dimension of the speaker embedding                                         \\ \hline
\end{tabular}
\end{table}

The initial learning rate is \(1e^{-3}\) and decays by 0.5 if the accuracy of validation set was not improved in 2 epochs. Early stopping was applied if the accuracy of validation set have not improved for 6 epochs. As the same in SpEx+~\cite{ge2020spex+}, We use the multi-task learning implementation for training with two objectives. We use the scale-invariant signal-to-distortion ratio~\cite{SISDR} (SISDR) as the loss output speech quality and a cross-entropy (CE) loss for speaker classification:
\begin{equation}
\begin{aligned}
  SISDR &= 10\log_{10}(\frac{{\vert\vert e_{target} \vert\vert}^2}{{\vert\vert e_{res} \vert\vert}^2}) \\ &=  10\log_{10}(\frac{{\vert\vert {\frac{\hat{s}^T s}{{\vert\vert s \vert\vert}^2} s} \vert\vert}^2}{{\vert\vert {{\frac{\hat{s}^T s}{{\vert\vert s \vert\vert}^2} s} - \hat{s}} \vert\vert}^2})
  \label{eq5}
\end{aligned}
\end{equation}
\begin{equation}
  \hat{s} = e_{target} + e_{res}
  \label{eq6}
\end{equation}
\begin{equation}
  \mathcal{L}_{\mathrm{SI}-\mathrm{SDR}}=-\left[(1-\alpha-\beta) SISDR_{s_1}+\alpha SISDR_{s_2}+\beta SISDR_{s_3}\right]
  \label{eq6}
\end{equation}
\begin{equation}
  \mathcal{L}_{\mathrm{CE}}=-\sum_{i=1}^{N_s} I_i \log \left(\sigma(W \cdot v)_i\right)
  \label{eqCE}
\end{equation}
where \(\hat{s}\) represent the estimated speech. \(s_1\), \(s_2\) and \(s_3\) represent three different multi-scale estimation, respectively. \(\alpha\) and \(\beta\) are the weights to different scales.  \(e_{target}\) and \(e_{res}\) indicate the estimated speech's orthogonal projection and residual w.r.t. the reference speech, respectively. \(N_s\) is the number of speakers in the training datasets. \(W\) represents a weight matrix, \(\sigma(\cdot)\) represents a softmax function. The multi-task objective function for single speaker is defined as:

\begin{equation}
  \mathcal{L}_{\mathrm{multi}} = \mathcal{L}_{\mathrm{SI}-\mathrm{SDR}} + \mathcal{L}_{\mathrm{CE}}
  \label{eqmulti}
\end{equation}
The overall objective function for our MTSS model is defined as:
\begin{equation}
\begin{aligned}
  \mathcal{L}({\theta}|m,aux_{s_{1,2}},spk_{1,2},I_{s_{1,2}}) = {\lambda}_1\mathcal{L}_{\mathrm{multi_{spk1}}} +  {\lambda}_2\mathcal{L}_{\mathrm{multi_{spk2}}}
  \label{eqall}
\end{aligned}
\end{equation}
%

where \(\boldsymbol{m}\) is the input sequence, \(\boldsymbol{aux_{s_{1,2}}}\) are the reference speeches of two speakers, \({spk_{1}}\) and \({spk_{2}}\) are the target speeches of two speakers, \({\lambda}_{1}\) and \({\lambda}_{2}\) are the weights of scale-invariant signal-to-distortion ratio (SI-SDR) loss and cross-entropy (CE) loss, respectively. In this work, we set \({\lambda}_1 = 0.5\) and \({\lambda}_2 = 0.25\) as the default values.

The SD module chooses the Adam and binary cross-entropy loss as the optimizer. The input chunk size is 16s, and the acoustic feature is 80-dim log Mel-filterbank energies with a frame length of 25ms and a frame shift of 10ms. The training details can be found in \cite{wang2021dku}.
The trainging steps of the SD modeules are as follow:

\begin{enumerate}
      \item Transfer the pre-trained speaker embedding model's parameters to the front-end model in the TSVAD model. Maintain the front-end model in a fixed state while focusing our training efforts on the back-end model.
      \item Subsequently, once the back-end model reaches convergence, we proceed to unfreeze the front-end model and embark on a joint training phase for the entire model, spanning an additional 10 epochs.
      \item In the final stage, we initiate fine-tuning of the model using the AliMeeting training set, extending this process over 200 epochs while employing a learning rate of \(1e^{-5}\).
  \end{enumerate}

The Diarization Error Rate (DER) of the single-channel SD module~\cite{wang2022} on the test set of Alimeeting are shown in the Table~\ref{tab:DER}. We use the offline model as the SD system in our proposed SD-MTSS model. The SD module has a 4.12\% DER on the evalution set of the Alimeeting dataset. 

\begin{table}[]
\caption{The DERs(\%) of the single-channel offline and online SD systems on AliMeeting Eval set. }
  \label{tab:DER}
  \centering
\begin{tabular}{c|cll|c}
\hline
\textbf{model} & \textbf{2-spk} & \textbf{3-spk} & \textbf{4-spk} & \textbf{Total} \\ \hline
offline        & 0.89           & 6.63           & 5.47           & 4.12           \\ \hline
online         & 1.90           & 8.36           & 12.12          & 8.14           \\ \hline
\end{tabular}
\end{table}

We evaluate our proposed models for two steps: 1) Examining the performance of MTSS on WSJ0-2mix-extr dataset. We train the MTSS model with a pre-trained model on the training set of WSJ0-2mix-extr. Then, we compare MTSS-Softmax and MTSS-ReLU in terms of SDRi, SI-SDR, and PESQ. 2)  Examining the performance of SD-MTSS system on Alimeeting. We compare SpEx+\footnotemark[2]\footnotetext[2]{\url{https://github.com/gemengtju/SpEx_Plus}} (implemented by ourself with using Libri-2mix dataset as training set) and SD-MTSS model in terms of spekaer dependent character error rate (CER)\cite{fu2021aishell}.

\begin{table}[t]
\caption{SDR(dB), SI-SDR(dB), and PESQ of separated speech using the MTSS method on the WSJ0-2mix-extr dataset. \textbf{N} indicates the number of outputs per inference. \(s_1\) and \(s_2\) indicate the different speaker of the mixture. \textbf{MTSS-ReLU}: Using ReLU as the activation function and do not impose constraints on masks. \textbf{MTSS-Softmax}: Using softmax function to limit the sum of masks to 1.}
  \label{tab:wsj}
  \centering
\begin{tabular}{c|c|cc|cc|cc}
\toprule
\multirow{2}{*}{\textbf{Methods}} & \multirow{2}{*}{\textbf{N}} & \multicolumn{2}{c|}{\textbf{SDR}}                      & \multicolumn{2}{c|}{\textbf{SI-SDR}}                            & \multicolumn{2}{c}{\textbf{PESQ}}                            \\ \cline{3-8} 
                                  &                             & s1                        & s2                         & s1                        & s2                         & s1                       & s2                       \\ \hline
Mixture                           & -                           & 2.60                      & -2.14                      & 2.50                      & -2.50                      & 2.31                     & 1.86                     \\ \hline
SpeakerBeam~\cite{delcroix2018single}                       & 1                           & 9.62                      & -                          & 9.22                      & -                          & 2.64                     & -                        \\
SBF-MTSAL-Concat~\cite{xu2019optimization}                  & 1                           & 11.39                     & -                          & 10.60                     & -                          & 2.77                     & -                        \\
TseNet~\cite{xu2019time}                            & 1                           & 15.24                     & -                          & 14.73                     & -                          & 3.14                     & -                        \\
SpEx~\cite{xu2020spex}                              & 1                           & 17.15                     & -                          & 16.68                     & -                          & 3.36                     & -                        \\
SpEx+~\cite{ge2020spex+}                             & 1                           & 18.54                     & -                          & 18.20                     & -                          & 3.49                     & -                        \\ \hline
Pre-trained model\footnotemark[2]\footnotetext[2]{\url{https://github.com/gemengtju/SpEx_Plus}}                 & 1                           & \multicolumn{1}{l}{18.15} & \multicolumn{1}{l|}{16.42} & \multicolumn{1}{l}{17.55} & \multicolumn{1}{l|}{15.89} & \multicolumn{1}{l}{3.44} & \multicolumn{1}{l}{3.28} \\
MTSS-ReLU                       & 2                           & 19.18                     & 17.29                      & 18.72                     & 16.84                      & 3.56                     & 3.39                     \\
MTSS-Softmax                         & 2                           & \textbf{19.92}                     & \textbf{17.42}                      & \textbf{19.54}                     & \textbf{16.99}                      & \textbf{3.62}                     & \textbf{3.41}                     \\ 
\bottomrule
\end{tabular}
\end{table}

\section{Results and Discussions}

\subsection{Results on WSJ0-2mix-extr}
The results of our proposed MTSS model and the baseline system is shown in Table~\ref{tab:wsj}. Since we used the same simulation test set as \cite{ge2020spex+} uses, we directly use the evaluation results of SpeakerBeam, SBF-MTSAL-Concat, TseNet, SpEx, and SpEx+ in \cite{ge2020spex+}. As shown in  Table~\ref{tab:wsj}, SpEx+~\cite{ge2020spex+} is the baseline which we implemented, and MTSS are the model we proposed. Our proposed MTSS model achieves significantly better results across all the metrics. The samples of separated audio are available at this link \footnotemark[3]\footnotetext[3]{\url{https://github.com/ZBang/SD-MTSS}}. 
Specifically, MTSS-Softmax outperforms SpEx+ with relative improvements of 7.4\% in terms of SDR, 7.3\% in terms of SI-SDR, and 3.2\% in terms of PESQ, respectively. In addition, we get better improvement on each speaker (\(s_1\), \(s_2\)) while extracting their target speech simultaneously. Comparing the results of MTSS-ReLU and MTSS-Softmax, we can conclude that setting the constraint for each speaker's mask mainly contribute to the improvements, and the interrelationship between different speakers at each frame benefits the model to extract the target source.

\begin{table}[]
\caption{Average \textbf{speaker dependent CER(\%)} results of SD-MTSS on Eval\_Ali\_far and Test\_Ali\_far sets. \textbf{N} indicates the number of outputs per inference. \textbf{Re} indicates that the model is implemented by ourself. We use MTSS-Softmax model as the MTSS module of the SD-MTSS system.}
\label{tab:ASR}
\centering
\begin{tabular}{c|c|c|c|c}
\hline
\textbf{Methods} & \textbf{N} & \multicolumn{1}{l|}{\textbf{Eval}} & \textbf{Test} & \textbf{Avg} \\ \hline
Mixture          & -          & 96.70                                            & 95.51                      & 95.83           \\
SpEx+~\cite{ge2020spex+} (Re)            & 1          & 45.80                                            & 43.79                       & 44.34            \\
SD-MTSS          & 2          & \textbf{35.97}                                   & \textbf{35.78}                       & \textbf{35.83}            \\ \hline
\end{tabular}
\begin{tablenotes}
\footnotesize
\item[1] * We use WeNet~\cite{yao2021wenet} as the speech recognition model in this experiments. The wenet model is trained by WeNetSpeech~\cite{9746682} and our inhouse data together with approximately 15K hours.
\end{tablenotes}
\end{table}

\subsection{Results on Alimeeting}
The speech recognition results of our proposed SD-MTSS system are shown in Table~\ref{tab:ASR}. Here, we report the average speaker independent CER results on the Eval\_Ali\_far and Test\_Ali\_far subsets of Alimeeting. It is important to note that we have adopted Minimum Variance Distortionless Response (MVDR\footnotemark[4]\footnotetext[4]{\url{https://github.com/funcwj/setk}}) beamformer on the mixture in advance. Due to multi-speaker interference, many insertion errors are generated in recognition of the mixed speech. The difference between SpEx+ and our proposed SD-MTSS system is that the SD-MTSS can extract the speech of each speaker simultaneously in one inference and does not need an enrollment wave in advance. Compared with the SpEx+ model, the SD-MTSS model achieves a 21.4\% and 18.3\% relative average speaker dependent CER reduction on Eval\_Ali\_far and Test\_Ali\_far subsets  of the Alimeeting, respectively. Since we only need to evaluate the effectiveness of SD-MTSS model, we did not train the recognition and separation models jointly. As far as we know, joint training and fine-tuning with Alimeeting datasets can improve final recognition performance~\cite{yu2022comparative}.  
The results of our proposed SD-MTSS system are shown in Table~\ref{tab:ASR}. Since we evaluate the system on the far-field data and use the corresponding close-talking data as the ground truth, the model does not performance well in terms of SDRi and SI-SDRi. Nevertheless, from Table~\ref{tab:ASR}, we can draw two conclusions: 1) Our proposed multiple target speech separation model surpasses the pre-trained model (SpEx+) with a large margin in terms of SI-SDRi. 2) Modifying the estimated speech with the TSVAD decision as shown in the Figure~\ref{fig:MTSS} could significantly enhance the system's performance, for this approach somehow solves the non-target speaker residual problem.

\section{Conclusions}
In this work, we propose a Multiple Target Speech Separation (MTSS) model which can simultaneously extract each speaker's voice from the mixed speech. To establish a relationship between different speakers in each frame, we constrain the sum of each speaker's estimated mask to 1 when extracting their speech simultaneously. Moreover, we proposed a speaker diarization multiple target speech separation system (SD-MTSS), which consists of a speaker diarization (SD) module and a MTSS module. By associating the speaker diarization task and the target speech separation task together, we do not require the additional reference speech for enrollment. The experimental results show that our proposed MTSS model significantly improves the separation performance on WSJ0-2mix-extr datasets. In addition, the SD-MTSS model outperforms the baseline by a large margin in terms of speaker independent CER on the Alimeeting datasets. For future works, we will implement our method with different state-of-art networks and improve the system's performance in the far-field scenarios.

\section{Acknowledgements}
This research is funded in part by the National Natural Science Foundation of China (62171207) and Science and Technology Program of Guangzhou City (202007030011) and OPPO. Many thanks for the computational resource provided by the Advanced Computing East China Sub-Center.
%
%
%
%

\end{document}